\documentclass[12pt,thmsa]{article}

\usepackage{cite}

\usepackage{graphicx}

\usepackage{dcolumn}

\begin{document}

\title{Rational approximation to the Thomas--Fermi equation}

\author{Francisco M. Fern\'{a}ndez \thanks{e--mail: fernande@quimica.unlp.edu.ar}\\
INIFTA (UNLP, CCT La Plata--CONICET), Divisi\'{o}n Qu\'{i}mica Te\'{o}rica,\\
Diag. 113 y 64 (S/N), Sucursal 4, Casilla de Correo 16,\\
1900 La Plata, Argentina}

\maketitle

\begin{abstract}
We discuss a recently proposed analytic solution to the Thomas--Fermi (TF)
equation and show that earlier approaches provide more accurate results.
In particular, we show that a simple and straightforward rational
approximation to the TF equation yields the slope at origin with unprecedented
accuracy, as well as remarkable values of the TF function and its first
derivative for other coordinate values.
\end{abstract}

\section{Introduction\label{sec:Intro}}

The Thomas--Fermi (TF) equation has proved useful for the treatment of many
physical phenomena that include atoms\cite{BCR74,CM50,M57,MT79,M83},
molecules\cite{M52,M57}, atoms in strong magnetic fields\cite{BCR74,MT79,M83}%
, crystals\cite{UT55} and dense plasmas\cite{YK89} among others. For that
reason there has been great interest in the accurate solution of that
equation, and, in particular, in the accurate calculation of the slope at
origin\cite{KMNU55,PP87,FO90}. Besides, the mathematical aspects of the TF
equation have been studied in detail\cite{H69,H70}. Some time ago Liao\cite
{L03} proposed the application of a technique called homotopy analysis
method (HAM) to the solution of the TF equation and stated that ``it is the
first time such an elegant and explicit analytic solution of the
Thomas--Fermi equation is given''. This claim is surprising because at first
sight earlier analytical approaches are apparently simpler and seem to have
produced much more accurate results\cite{PP87,FO90,T91,EFGP99}. Recently,
Khan and Xu\cite{KX07} improved Liao's HAM by the addition of adjustable
parameters that improve the convergence of the perturbation series.

The purpose of this paper is to compare the improved HAM with a
straightforward analytical procedure based on Pad\'{e} approximants\cite
{EFGP99} supplemented with a method developed some time ago\cite
{FMT89,F92,FG93,F95,F95b,F95c,F96,F96b,F97}. In Section \ref{sec:HAM} we
outline the main ideas of the HAM, in Section \ref{sec:HPM} apply the
Hankel--Pad\'{e} method (HPM) to the TF equation, and in Section \ref
{sec:conclusions} we compare the HAM with the HPM and with other approaches.

\section{The homotopy analysis method \label{sec:HAM}}

In order to facilitate later discussion we outline the main ideas behind the
application of the HAM to the TF equation. The TF equation
\begin{equation}
u^{\prime \prime }(x)=\sqrt{\frac{u(x)^{3}}{x}},\;u(0)=1,\;u(\infty )=0
\label{eq:TF}
\end{equation}
is an example of two--point nonlinear boundary--value problem. When solving
this ordinary differential equation one faces problem of the accurate
calculation of the slope at origin $u^{\prime }(0)$ that is consistent with
the physical boundary conditions indicated in equation (\ref{eq:TF}).

In what follows we choose the notation of Khan and Xu\cite{KX07} whose
approach is more general than the one proposed earlier by Liao\cite{L03}.
They define the new solution $g(\xi )=\gamma u(x)$, where $\xi =1+\lambda x$
and rewrite the TF equation as
\begin{equation}
(\xi -1)\lambda ^{3}\gamma g^{\prime \prime }(\xi )^{3}-g(\xi )^{3}=0
\label{eq:TF2}
\end{equation}
where $\gamma $ is the inverse of the slope at origin ($u^{\prime
}(0)=1/\gamma $) and $\lambda $ is an adjustable parameter. Khan and Xu\cite
{KX07} state that the solution to Eq. (\ref{eq:TF2}) can be written in the
form
\begin{equation}
g(\xi )=\sum_{j=1}^{\infty }A_{j}\xi ^{-j}  \label{eq:g_series}
\end{equation}
that reduces to Liao's expansion\cite{KX07} when $\lambda =1$.

In principle there is no reason to assume that the series (\ref{eq:g_series}%
) converges and no proof is given in that sense\cite{L03,KX07}. Besides, the
partial sums of the series (\ref{eq:g_series}) will not give the correct
asymptotic behaviour at infinity\cite{H69,H70,BO78} as other expansions do%
\cite{PP87,FO90}.

Liao\cite{L03} and Kahn and Xu\cite{KX07} do not use the ansatz (\ref
{eq:g_series}) directly to solve the problem but resort to perturbation
theory. For example, Kahn and Xu\cite{KX07} base their approach on the
modified equation
\begin{equation}
(1-q)\mathcal{L}\left[ \Phi (\xi ;q)-g_{0}(\xi )\right] =q\hbar \mathcal{N}%
\left[ \Phi (\xi ;q),\Gamma (q)\right]  \label{eq:HAM}
\end{equation}
where $\mathcal{L}$ and $\mathcal{N}$ are linear and nonlinear operators,
respectively, $0\leq q\leq 1$ is a perturbation parameter and $\hbar $ is
another adjustable parameter. Besides, $g_{0}(\xi )$ is a conveniently
chosen initial function and $\Phi (\xi ;q)$ becomes the solution to equation
(\ref{eq:TF2}) when $q=1$\cite{KX07}. Both $\Phi (\xi ;q)$ and $\Gamma (q)$
are expanded in a Taylor series about $q=0$ as in standard perturbation
theory, and $\Gamma (0)=\gamma _{0}$ is another adjustable parameter\cite
{KX07}.

The authors state that HAM is a very flexible approach that enables one to
choose the linear operator and the initial solution freely\cite{L03,KX07}
and also to introduce several adjustable parameters\cite{KX07}. However, one
is surprised that with so many adjustable parameters the results are far
from impressive, even at remarkable great perturbation orders\cite{L03,KX07}%
. For example the $[30/30]$ Pad\'{e} approximant of the HAM series yields $%
u^{\prime }(0)$ with three exact digits\cite{KX07}, while the $[1/1]$
Pad\'{e} approximant of the $\delta $ expansion\cite{BMPS89} provides
slightly better results\cite{L90,C93}. A more convenient expansion of the
solution of the TF equation leads to many more accurate digits\cite
{PP87,FO90} with less terms.

\section{The Hankel--Pad\'{e} method \label{sec:HPM}}

In what follows we outline a simple, straightforward analytical method for
the accurate calculation of $u^{\prime }(0)$. In order to facilitate the
application of the HPM we define the variables $t=x^{1/2}$ and $%
f(t)=u(t^{2})^{1/2}$, so that the TF equation becomes
\begin{equation}
T(f,t)=t\left[ f(t)f^{\prime \prime }(t)+f^{\prime }(t)^{2}\right]
-f(t)f^{\prime }(t)-2t^{2}f(t)^{3}=0  \label{eq:TF3}
\end{equation}
We expand the solution $f(t)$ to this differential equation in a Taylor
series about $t=0$:
\begin{equation}
f(t)=\sum_{j=0}^{\infty }f_{j}t^{j}  \label{eq:f_series}
\end{equation}
where the coefficients $f_{j}$ depend on $f_{2}=f^{\prime \prime
}(0)/2=u^{\prime }(0)/2$. On substitution of the series (\ref{eq:f_series})
into equation (\ref{eq:TF3}) we easily calculate as many coefficients $f_{j}$
as desired; for example, the first of them are
\begin{equation}
f_{0}=1,\;f_{1}=0,\;f_{3}=\frac{2}{3},\;f_{4}=-\frac{f_{2}^{2}}{2},\;f_{5}=-%
\frac{4f_{2}}{15},\ldots
\end{equation}

The HPM is based on the transformation of the power series (\ref{eq:f_series}%
) into a rational function or Pad\'{e} approximant
\begin{equation}
\lbrack M/N](t)=\frac{\sum_{j=0}^{M}a_{j}t^{j}}{\sum_{j=0}^{N}b_{j}t^{j}}
\label{eq:[M/N]}
\end{equation}
One would expect that $M<N$ in order to have the correct limit at infinity;
however, in order to obtain an accurate value of $f_{2}$ it is more
convenient to choose $M=N+d$, $d=0,1,\ldots $ as in previous applications of
the approach to the Schr\"{o}dinger equation (in this case it was called
Riccati--Pad\'{e} method (RPM))\cite
{FMT89,F92,FG93,F95,F95b,F95c,F96,F96b,F97}.

The rational function (\ref{eq:[M/N]}) has $2N+d+1$ coefficients that we may
choose so that $T([M/N],t)=\mathcal{O}(t^{2N+d+1})$ and the coefficient $%
f_{2}$ remains undetermined. If we require that $T([M/N],t)=\mathcal{O}%
(t^{2N+d+2})$ we have another equation from which we obtain $f_{2}$.
However, it is convenient to proceed in a different (and entirely
equivalent) way and require that
\begin{equation}
\lbrack M/N](t)-\sum_{j=0}^{2N+d+1}f_{j}t^{j}=\mathcal{O}(t^{2N+d+2})
\label{eq:[M/N]2}
\end{equation}
In order to satisfy this condition it is necessary that the Hankel
determinant vanishes
\begin{equation}
H_{D}^{d}=\left| f_{i+j+d+1}\right| _{i,j=0,1,\ldots N}=0,  \label{eq:Hankel}
\end{equation}
where $D=N+1$ is the dimension of the Hankel matrix. Each Hankel determinant
is a polynomial function of $f_{2}$ and we expect that there is a sequence
of roots $f_{2}^{[D,d]}$, $D=2,3,\ldots $ that converges towards the actual
value of $u^{\prime }(0)/2$ for a given value of $d$. We compare sequences
with different values of $d$ for inner consistency (all of them should give
the same limit). Notice that a somewhat similar idea was also proposed by Tu%
\cite{T91}, although he did not develop it consistently.

Present approach is simple and straightforward: we just obtain the Taylor
coefficients $f_{j}$ from the differential equation (\ref{eq:TF3}) in terms
of $f_{2}$, derive the Hankel determinant, and calculate its roots. Since $%
f_{4}$ is the first nonzero coefficient that depends on $f_{2}$ we choose
Hankel sequences with $d\geq 3$.

The Hankel determinant $H_{D}^{d}$ exhibits many roots and their number
increases with $D$. If we compare the roots of $H_{D}^{d}$ with those of $%
H_{D-1}^{d}$ we easily identify the sequence $f_{2}^{[D,d]}$ that converges
towards the actual value of $f_{2}$. Fig. \ref{fig:logconv} shows $\log
\left| 2f_{2}^{[D,d]}-2f_{2}^{[D-1,d]}\right| $ for $D=3,4,\ldots $ that
provides a reasonable indication of the convergence of the sequence of
roots. We clearly appreciate the great convergence rate of the sequences
with $d=3$ and $d=4$. For example, for $d=3$ and $D\leq 30$ it is
approximately given by
$\left| 2f_{2}^{[D,3]}-2f_{2}^{[D-1,3]}\right|=14.2\times 10^{
-0.705D}$. From the sequences for $D\leq 30$ we estimate
$u^{\prime }(0)=-1.58807102261137531$ which we believe is accurate to the
last digit. We are not aware of a result of such accuracy in the literature
with which we can compare our estimate. It is certainly far more accurate
than the result obtained by Kobayashi et al\cite{KMNU55} by numerical
integration that is commonly chosen as a benchmark\cite{L03,KX07}.

Present rational approximation to the TF function is completely different
from previous application of the Pad\'{e} approximants, where the slope at
origin was determined by the asymptotic behaviour of at infinity\cite{EFGP99}%
. Our approach applies to $u(x)^{1/2}$ and the slope at origin is determined
by a local condition at that point (\ref{eq:[M/N]2}) which results in the
Hankel determinant (\ref{eq:Hankel}). In this sense our approach is similar
to (although more systematic and consistent than) Tu's one\cite{T91} as
mentioned above.

Once we have the slope at origin we easily obtain an analytical expression
for $u(x)$ in terms of the rational approximation (\ref{eq:[M/N]}) to $f(t)$%
. In order to have the correct behaviour at infinity we choose $N=M+3$\cite
{EFGP99}. Table~\ref{tab:u(x)} shows values of $u(x)$ and its first
derivative for $1<x<1000$ (the approximation is obviously much better for $%
0<x<1$) given by the approximant $[5/8]$. Our results are in remarkably
agreement with the numerical calculation of Kobayashi et al\cite{KMNU55} and
are by far much more accurate than those provided by the HAM\cite{L03,KX07}.
Notice that we are comparing a $[5/8]$ Pad\'{e} approximant on the
straightforward series expansion (\ref{eq:f_series}) with $[50/50]$ and $%
[30/30]$ approximants on an elaborated perturbation series\cite{L03,KX07}.

\section{Conclusions \label{sec:conclusions}}

\begin{figure}[H]
\begin{center}
\includegraphics[width=9cm]{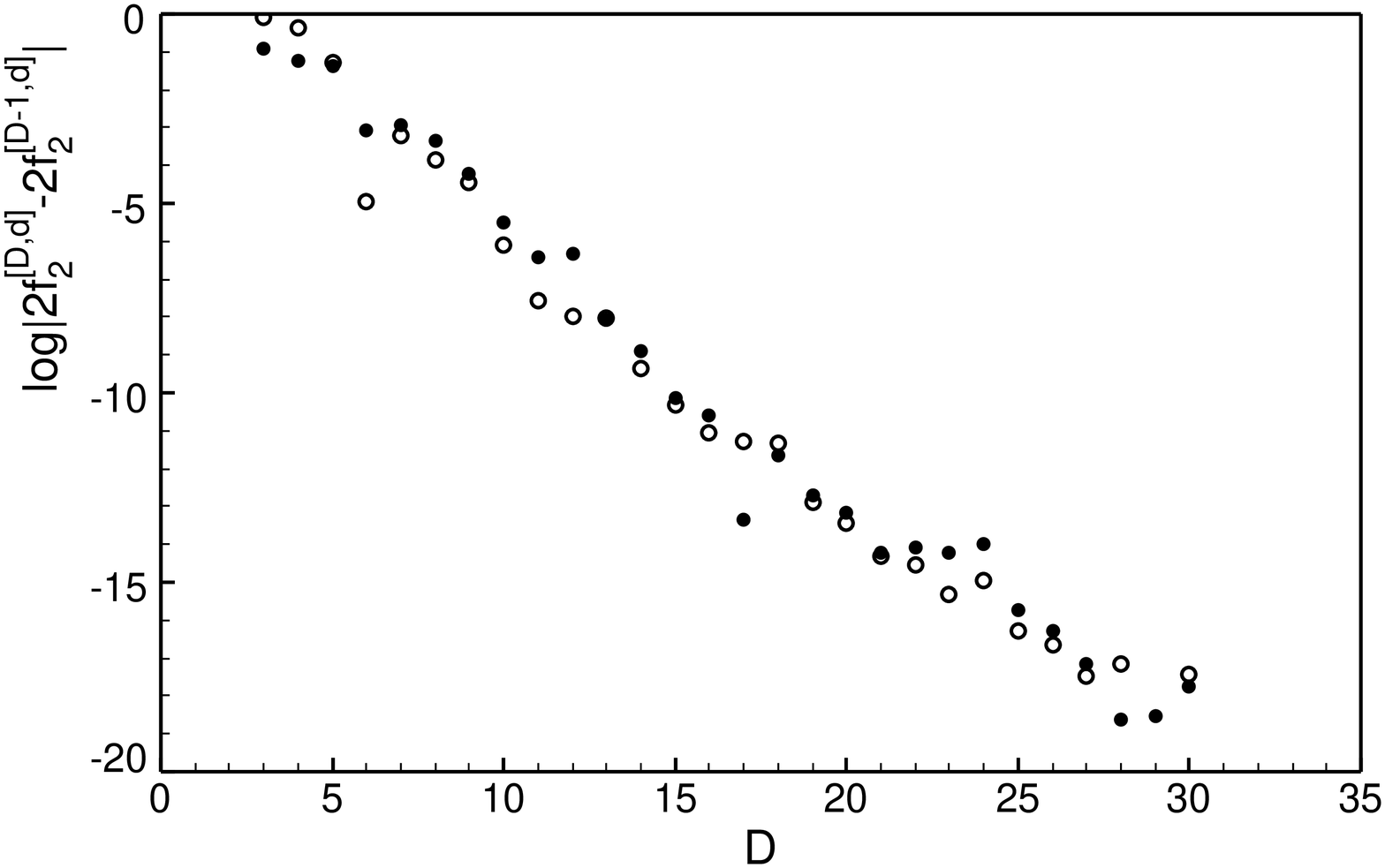}
\end{center}
\caption{$\log \left| 2f_{2}^{[D,d]}-2f_{2}^{[D-1,d]}\right| $ for $d=3$
(circles) and $d=4$ (filled circles) }
\label{fig:logconv}
\end{figure}

Any accurate analytical expression of the solution $u(x)$ to the TF equation
requires an accurate value of the unknown slope at origin $u^{\prime }(0)$,
and the HPM provides it in a simple and straightforward way. In this sense
the HPM appears to be preferable to other accurate approaches\cite
{PP87,FO90,KMNU55} and is far superior to the HAM\cite{L03,KX07}. Notice for
example that our estimate $2f_{2}^{[5,3]}=-1.588$, based on a rational
approximation $[7/4]$, is better than the result provided by a $[30/30]$ Pad%
\'{e} approximant on the improved HAM perturbation series\cite{KX07}.
Besides, by comparing Table 2 of Khan and Xu\cite{KX07} with our Fig. \ref
{fig:logconv} one realizes the different convergence rate of both
approaches. One should also take into account that the HPM does not have any
adjustable parameter for tuning up its convergence properties, while, on the
other hand, the ``flexible'' HAM with some such parameters plus a Pad\'{e}
summation results in a much smaller convergence rate\cite{L03,KX07}.

We also constructed a Pad\'{e} approximant $[5/8]$ from the series (\ref
{eq:f_series}) and obtained the TF function and its derivative with an
accuracy that outperforms the $[50/50]$ and $[30/30]$ Pad\'{e} approximants
on the HAM perturbation series\cite{L03,KX07}. It is clear that the HPM is
by far simpler, more straightforward, and much more accurate than the HAM.

In addition to the physical utility of the HPM we think that its
mathematical features are most interesting. Although we cannot provide a
rigorous proof of the existence of a convergent sequence of roots for each
nonlinear problem, or that the sequences will converge towards the correct
physical value of the unknown, a great number of successful applications to
the Schr\"{o}dinger equation\cite{FMT89,F92,FG93,F95,F95b,F95c,F96,F96b,F97}
suggest that the HPM is worth further investigation. Notice that we obtain a
global property of the TF equation $u^{\prime }(0)$ from a local approach:
the series expansion about the origin (\ref{eq:f_series}). The fact that our
original rational approximation (\ref{eq:[M/N]}) does not have the correct
behaviour at infinity is not at all a  problem because we may resort to a
more conventient expansion\cite{EFGP99} once we have an accurate value of
the unknown slope at origin.

Finally, we mention that the HPM has recently proved successful for the
treatment of other two--point nonlinear equations\cite{AF07} of interest in
some fields of physics\cite{BFG07,BBG08,BBG08b}.

\begin{table}[H]
\caption{Values of the Thomas-Fermi function and its derivative}
\label{tab:u(x)}
\begin{center}
\par
\begin{tabular}{D{.}{.}{2}D{.}{.}{12}D{.}{.}{12}}
\hline
\multicolumn{1}{c}{$x$}& \multicolumn{1}{c}{$u(x)$}&
\multicolumn{1}{c}{$-u^\prime(x)$} \\
\hline
 1& 0.424008   &   0.273989   \\
 5& 0.078808   &   0.023560  \\
10& 0.024315   &   0.0046028 \\
20& 0.005786   &   0.00064727 \\
30& 0.002257& 0.00018069      \\
40& 0.001114& 0.00006969      \\
50& 0.000633& 0.00003251      \\
60& 0.000394& 0.0000172       \\
70& 0.0002626&0.000009964     \\
80& 0.0001838& 0.000006172    \\
90& 0.0001338& 0.000004029    \\
100&0.0001005& 0.000002743    \\
1000&0.000000137&0.00000000040\\

\hline

\end{tabular}
\par
\end{center}
\end{table}

\end{document}